\documentclass[review]{elsarticle}
\usepackage[top=3.5cm,bottom=3.5cm,left=3.5cm,right=3.5cm]{geometry}
\usepackage{hyperref}
\hypersetup{colorlinks = true,linkcolor = blue,anchorcolor =red,citecolor = blue,filecolor = red,urlcolor = red,
            pdfauthor=author}
\usepackage{amsmath}
\usepackage{amssymb}
\usepackage{supertabular,booktabs}
\usepackage{float}
\usepackage[table]{xcolor}
\definecolor{shade1}{RGB}{255,255,204}
\definecolor{shade2}{RGB}{255,237,160}
\definecolor{shade3}{RGB}{254,217,118}
\definecolor{shade4}{RGB}{254,178,76}
\definecolor{shade5}{RGB}{253,141,60}
\definecolor{shade6}{RGB}{252,78,42}
\definecolor{shade7}{RGB}{227,26,28}
\definecolor{shade8}{RGB}{177,0,38}

\usepackage{url}

\journal{---} 

\begin{document}

\begin{frontmatter}

\title{Large-Scale Scenarios of Electric Vehicle Charging with a Data-Driven Model of Control}

\author[mythirdaddress]{Siobhan Powell\corref{mycorrespondingauthor}}
\cortext[mycorrespondingauthor]{Corresponding author}
\ead{siobhan.powell@stanford.edu}
\ead[url]{s3l.stanford.edu}
\ead[url]{gismo.slac.stanford.edu}
\author[mysecondaryaddress,mymainaddress]{Gustavo Vianna Cezar}
\author[mysecondaryaddress]{Elpiniki Apostolaki-Iosifidou}
\author[mymainaddress]{Ram Rajagopal\corref{mycorrespondingauthor}}
\ead{ramr@stanford.edu}

\address[mythirdaddress]{Department of Mechanical Engineering, Stanford University}
\address[mysecondaryaddress]{Applied Energy Division, SLAC National Accelerator Laboratory}
\address[mymainaddress]{Department of Civil and Environmental Engineering, Stanford University}

\begin{abstract}
Planning to support widespread transportation electrification depends on detailed estimates for the electricity demand from electric vehicles in both uncontrolled and controlled or smart charging scenarios. We present a modeling approach to rapidly generate charging estimates that include control for large-scale scenarios with millions of individual drivers. We model uncontrolled charging demand using statistical representations of real charging sessions. We model the effect of load modulation control on aggregate charging profiles with a novel machine learning approach that replaces traditional optimization approaches. We demonstrate its performance modeling workplace charging control with multiple electricity rate schedules, achieving small errors (2.5\% to 4.5\%), while accelerating computations by more than 4000 times. We illustrate the methodology by generating scenarios for California's 2030 charging demand including multiple charging segments and controls, with scenarios run locally in under 50 seconds, and for assisting rate design modeling the large-scale impact of a new workplace charging rate. 
\end{abstract}


\end{frontmatter}

\section{Introduction} \label{sec:intro}

The transportation sector is undergoing a dramatic change. By 2040, nearly 500 million passenger plug-in electric vehicles (EVs) are expected to be on the road worldwide; a 71$\times$ increase from today \cite{bnef2020}. More than 300 million chargers split between workplace, public and residential locations will be deployed to support them. Charging decisions couple the electric power grid and transportation network, but the timing, location and power rate of future charging events are difficult to predict, depending on drivers' needs and behaviour. 

Long term planning in both the grid and transportation sectors requires estimates of the future power demand for charging in regions with millions of EVs. Due to the intrinsic uncertainty in the factors that drive EV use, planners need to consider many scenarios. Models utilized to generate these scenarios must be data-driven and scalable to millions of drivers. Moreover, charging control is critical for supporting high levels of EV adoption \cite{crozier2020opportunity, szinai2020reduced, brinkel2020should}, and its impact on millions of drivers needs to be estimated. 

Existing models for uncontrolled charging power demand can be broadly categorized as trip-based and charging data-based approaches. Trip-based approaches utilize detailed travel surveys or cellphone GPS data to simulate the travel patterns of individual drivers and determine their charging behavior \cite{wood2018new, sheppard2017modeling, e3_2020}. Statistical methods have also been used for this purpose, including to extend travel data to cover neighboring regions \cite{li2018gis}, simulate trip chains from origin-destination matrices \cite{mureddu2018complex, mu2014spatial}, fit distributions of trip parameters \cite{tang2015probabilistic, wang2017modeling}, use clustering to identify patterns in driver travel \cite{xu2018planning, crozier2019stochastic, sodenkamp2019can} and generate new trips by modeling vehicle location and state transitions as stochastic processes \cite{ul2018probabilistic, wang2018markov}. Charging data-based models utilize charging sessions data to directly learn charging demand profiles \cite{sadeghianpourhamami2018quantitive, nicholas2017lessons, flammini2019statistical, morrissey2016future, quiros2018statistical, brady2016modelling, dias2018impact, neaimeh2015probabilistic}. Statistical models are used to fit charging session parameters from existing data sets, and then can be utilized to sample profiles accordingly \cite{morrissey2016future, quiros2018statistical, flammini2019statistical, dias2018impact, brady2016modelling}. 
 
 There are three common types of controlled charging: load shifting, which shifts the charging later within the session time window; load modulation, which modulates the charging rate throughout the session to reshape the load profile; and driver-focused control, which uses mechanisms like pricing to influence drivers' charging location or timing decisions. For residential, workplace, or slow public charging, where drivers require only that their charging be complete before their departure, all three can apply. Fast public charging can only utilize driver-focused control. In this paper we focus on simulating control impacts for workplace and residential charging. 
 
 Workplace charging is increasingly common in the US as it benefits commuters and the grid \cite{wu2018role, chakraborty2019demand, mclaren2016co2}. Controlling workplace charging using load modulation brings numerous benefits including reduced costs, increased equipment lifetime and potential for grid services \cite{zhang2018value, kara2015estimating, powell2020controlled}. There is a vast literature on  algorithms for realistic control implementation (e.g.  \cite{lee2016adaptive}). Similar to workplace charging, residential charging can be controlled by load shifting and load modulation \cite{dias2018impact, coignard2019will, mobarak2019vehicle, crozier2020opportunity}. 
 
 In both applications, the existing algorithms for load modulation require solving optimization problems that do not scale for large urban-scale simulations. Generally such algorithms are too computationally expensive since they require tracking multiple constraints for each individual vehicle in the simulation. Approximating EV aggregates \cite{tang2016aggregated} addresses this problem at the expense of capturing necessary real world constraints.

  This paper proposes a fast, scalable, data-driven approach to estimate the aggregate load profile for large-scale EV charging scenarios for long term planning under different input assumptions, including estimates of controlled workplace and residential charging. The aggregate charging load at a given location with sufficient chargers is consistent day to day despite variation in the individual vehicle sessions. We use this insight to propose a methodology that learns, directly from data, how load modulation control reshapes the aggregate at any given location. The resulting model is used to present a range of scenarios for California's future charging load with a range of control objectives. The code to run the model is published in an open-source tool \cite{cezar2021}.

 The remainder of the paper is outlined as follows: in Section \ref{sec:model} we introduce the overall methodology; in Section \ref{sec: gmm} we detail the sessions model; in Section \ref{sec:control} we detail the control model; in Section \ref{sec:experiments} we apply it to a large data set of real charging sessions and we demonstrate the control model for workplace charging; in Section \ref{sec:results} we use the simulation approach to study scenarios for California's 2030 charging demand and we illustrate its application in rate design; in Section \ref{sec:discussion} we discuss; and in Section \ref{sec:conclusion} we conclude the study.

\section{Modeling Approach} \label{sec:model}

\begin{figure*}
    \centering
    \includegraphics[width=\linewidth]{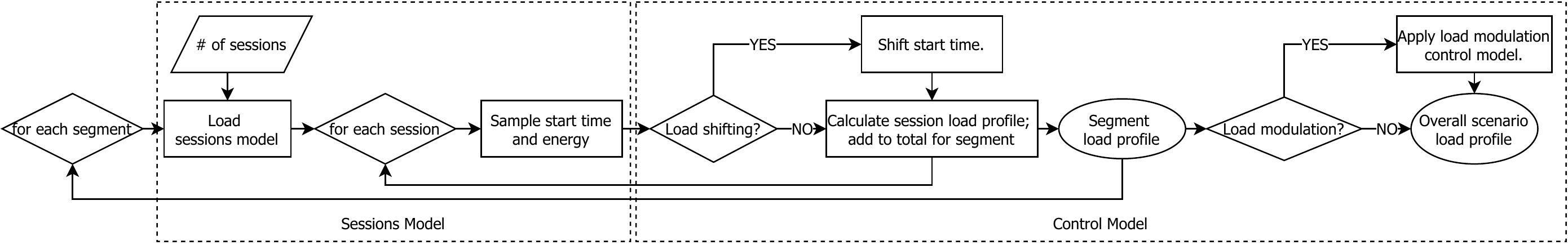}
    \caption{The algorithm used to generate the load profile for a given scenario. Each segment modeled with this approach is defined by its location, charging rate, and whether the model is for a weekday or weekend. The number of sessions in each segment is a user input. The control model and the sessions models are learned in advance offline and applied here to generate new scenarios. }
    \label{fig:flowchart}
\end{figure*}

Our modeling approach consists of two main components: a sessions model for uncontrolled charging and a control model. Both are built using a large data set of charging sessions.  Figure \ref{fig:flowchart} illustrates the algorithm followed to generate a new scenario using our methodology, highlighting the contributions of the sessions and control model components.  The sessions model uses statistical mixture models to capture the distributions of uncontrolled charging session parameters. Load modulation control is implemented in the control model by using a machine learning approach to learn the load shape impact of applying control.

Our approach divides the charging load into segments by charging speed and location, separating single family residences, multi-unit dwellings (MUDs), workplaces, and public charging. These segments are all common in present day charging patterns, and each presents unique usage and load patterns. In the United States, charging speed is categorized into levels one (L1), two (L2), and three (L3). Level three is also known as direct current fast charging (DCFC). For single family residences we consider both L1 and L2, for multi-unit dwellings and workplaces we include only L2, based on our charging data, and for public charging we consider both L2 and DCFC. L1 is wall-plug charging at 1.4 kW, L2 occurs at 6.6 kW in our data set, and we model DCFC at 150 kW. Modelers with other use-cases could tailor the approach by adjusting the segments, for example by conditioning on geographic location or vehicle make. 

The number of drivers in the simulation is an exogeneous parameter. The process shown in Figure \ref{fig:flowchart} takes the number of sessions in each segment as an input. This is calculated in our scenarios using two key assumptions: the drivers' frequency of charging, and the split between drivers who depend on charging in each segment.
Stepping through the flow chart, the sessions models generate parameters for each session in each segment, the control model can be applied, either to the individual sessions for load shifting control or to the aggregated sessions for load modulation control, and the sum of all sessions and segments returns the overall scenario load profile. 

In both the sessions and control components our approach depends on fitting the models in advance so that new scenarios can be generated based purely on the model objects, without needing to either run new optimizations or sample from the raw data directly.
This design improves both speed and privacy. Once each component model is learned, it is computationally inexpensive to generate new profiles. This allows the modeler to adjust assumptions and apply control rules very quickly. Separating the two steps of training and applying the models also means that no individual user's data is used when generating scenarios and the model can be stored and published without sharing sensitive data.

\section{Sessions Model} \label{sec: gmm}
We fit a separate sessions model to the subset of charging data corresponding to each segment. Since the segments are conditioned on charging rate and we assume the segments are initially uncontrolled, the two variables required to define each session are the start time and the energy. 
For each session, $i$, we define the variable, $x_i$, as 
\begin{equation}
    x_i = \begin{bmatrix} s_i \\ e_i \end{bmatrix} \text{,}
\end{equation}
where $s_i$ denotes the session's start time and $e_i$ denotes the energy delivered to the vehicle.

Coupling these two by modeling their joint distribution captures important differences in driver behaviour throughout the day. For example, workplace charging sessions which start in the afternoon tend to deliver lower energies than those which start during the morning peak. Studies using other statistical methods have also noted the importance of capturing correlations and joint dynamics between charging parameters \cite{brady2016modelling}. 

In \cite{quiros2018statistical} the authors demonstrated the efficacy of Gaussian Mixture Models (GMMs) to capture the distributions of charging parameters. We extend that approach to model the joint distribution of the parameters rather than modeling them independently.

A Gaussian Mixture Model comprises $G$ component multivariate Gaussian distributions \cite{mcnicholas2017mixture}. Each component, $g$, is described by mean vector, $\mathbf{\mu}_g$, and covariance matrix, $\Sigma_g$, and the component is weighted in the mixture according to a weight, $\pi_g$. 
When applied to a set of $p$-dimensional sessions data, $x_1, \hdots, x_n$, for analysis, the likelihood of the model can be expressed as 
\begin{equation}
    \mathcal{L} = \prod_{i=1}^n\sum_{g=1}^G\pi_g\phi(x_i|\mu_g,\Sigma_g) \text{,}
\end{equation}
where $\phi(x_i|\mu_g, \Sigma_g)$ denotes the density of $x_i$ in component multivariate Gaussian distribution, $g$.

The model parameters, $\pi_g$, $\mu_g$, and $\Sigma_g$, are estimated by applying an expectation-maximization algorithm to the log likelihood \cite{mcnicholas2017mixture}. In our application, since $x_i$ is a vector with $2$ components, each $\pi_g$ and $\mu_g$ is a vector in $\mathbb{R}^2$ and $\Sigma_g$ is a matrix in $\mathbb{R}^{2\times 2}$. We used the Bayesian Information Criterion (BIC) to select the number of components, $G$, and allowed that parameter to vary between the models for different segments \cite{claeskens2018model}. 

In addition to providing a statistical representation of the data, the Gaussian Mixture Models can be used as a tool to generate new sessions consistent with the distributions. Sampling from each Gaussian Mixture Model in the model gives us a sessions vector, $x$, of arrival time and energy consistent with the original distribution.
After fitting and saving the Gaussian Mixture Model for each segment in the data, we use them to sample in this way to generate new sessions, as shown in our load profile algorithm in Figure \ref{fig:flowchart}.

\section{Control Model} \label{sec:control}
In this paper we model two kinds of controlled charging: load shifting and load modulation. 

We refer to load shifting as `timer control' because it is most often implemented through the use of timers, either in the vehicle or charging station, which delay the start of charging to a certain time. Timer control has been included in other load modeling tools; we build on their work by including them here. 
Load modulation involves reshaping the charging load by modulating the vehicle's charging rate throughout its session. We present a novel approach using machine learning to represent load modulation control. 

\subsection{Load Shifting Control}
Timer control can be implemented by artificially changing a session start time in the step shown in Figure \ref{fig:flowchart}. A given session, $x_i$, can be altered to include timer control, $x_i'$:
\begin{equation}
    x_i = \begin{bmatrix} s_i \\ e_i \end{bmatrix} \longrightarrow x_i' = \begin{bmatrix} S \\ e_i \end{bmatrix} \text{,}
\end{equation}
so long as the timer start time, $S$, is within the window the vehicle is plugged in. The window between $S$ and the vehicle's departure should give enough time for the vehicle's energy requirement, $e_i$, to be satisfied. This is easily satisfied for overnight residential L2 charging sessions and many utility electricity rates set pricing that encourages drivers to delay their charging until after the evening peak load.

\subsection{Load Modulation Control}
We present a novel data-driven approach for modeling the change in the load shape under load modulation controlled charging.

The use of data-driven models to replace more expensive calculations is a common application for machine learning. In electricity systems, example applications include calculations of optimal power flow and models of the distribution grid \cite{yu2017robust, yu2017patopa, liu2018data} and voltage regulation \cite{chen2020data}. With this paper we bring that approach to a new application: to our knowledge no other work has explored data-driven modeling of large-scale controlled charging.

The charging control implemented by an aggregator defines an optimization problem: minimize costs subject to constraints defined by each vehicle's arrival, departure, and energy needs. That optimization problem can be viewed as a change to the uncontrolled aggregate charging load. For a given instance, $j$, on one day in one parking lot or one neighborhood, let $X_j$ represent the uncontrolled load and $Y_j$ represent the controlled load, the output of running the optimization. $X_j$ and $Y_j$ are both vectors of length $T$. Then,
\begin{equation}
    Y_j = f(X_j)
\end{equation}
represents the mapping implemented by the controlled charging optimization. 

We propose a data-driven approach to modeling $f$ to estimate the shape change. 
To build the input and output data for our machine learning model we must generate a large number of input-output pairs, $(X_j, Y_j)$. We do this by repeatedly sampling new instances of uncontrolled load, $X_j$, and running the charging control optimization defined above to calculate $Y_j$. Each instance is comprised of $n$ individual charging sessions sampled from the data. 
This mapping is depicted in Figure \ref{fig:traintest}.

\subsubsection{Optimization} \label{sec:optimization_definition}
Here we detail the optimization problem used to construct the training data, $X$ and $Y$.

Consider an individual site or area with $n$ vehicles under the control of an aggregator. Let $t$ index time, discretized into $T$ steps of size $\Delta t$. Let $L_t$ represent the total load of all vehicles at time $t$ measured in kilowatts. On a given day, vehicle $i$ arrives at time $t_{a, i}$, stays until time $t_{d, i}$, charges at rate $r_{i, t}$ throughout the day, and requires energy $e_i$ before departure. We define the total load as the sum over the individual vehicles at the site, 
\begin{equation}
    L_t = \sum_{i=1}^n r_{i, t} \text{,} \label{eq:l}
\end{equation}
and we define the constraints on $r$ imposed for each individual vehicle, $i \in \{1, \hdots n\}$:
\begin{align}
    r_{i, t} &= 0, ~~~~ t < t_{a, i} \label{eq:arr} \text{,}\\
    r_{i, t} &= 0, ~~~~ t \geq t_{d, i} \label{eq:dep}  \text{,}\\
    e_i &= \sum_{t=1}^T r_{i, t} \Delta t \text{.} \label{eq:ene}
\end{align}
These constraints ensure that charging cannot occur before arrival (Equation \ref{eq:arr}) or after departure (Equation \ref{eq:dep}), and that the session must deliver the required amount of energy (Equation \ref{eq:ene}). We constrain the charging rate to the minimum of the Level 2 charging rate and the individual vehicle's maximum charging rate, as some older vehicles have lower limits:
\begin{equation}
    r_{i, t} \leq r_{\text{max}, i}, ~~~~ \forall t \text{.} \label{eq:rmax}
\end{equation}
We also assume there is no vehicle to grid or bidirectional charging in this problem, so we enforce
\begin{equation}
    r \geq 0 \label{eq:rmin}
\end{equation}
for all vehicles at all times.

The aggregator's most common objective is to minimize the cost of supplying electricity. For residential charging, this is most commonly implemented by aligning with the lowest price period for energy. For workplace charging, most workplaces pay for electricity at business rates, including both energy and demand charge components, and in some parts of California there are rate plans specific to businesses providing workplace EV charging. A demand charge of $p_d$ dollars per kilowatt which applies to load within the time period $I$ has cost: 
\begin{equation}
    C = p_d \max_{t \in I} \left( L_t \right) \label{eq:dc} \text{.}
\end{equation}
An energy charge with price $p_e$ dollars per kilowatt hour in time period $I$ adds cost:
\begin{equation}
    C = p_e \Delta_t \sum_{t \in I} L_t \text{,} \label{eq:ec}
\end{equation}
where $\Delta_t$ is measured in hours. 
A rate schedule may have multiple energy and demand charge components applied at different times of day. We assume the optimization day will determine the month's demand charge. The total cost function for the optimization is the sum of those charges. In Section \ref{sec:experiments}, for comparison, we also implement a control scheme where the objective is directly to minimize the peak load: 
\begin{equation}
    C = \max_t L_t \text{.} \label{eq:minpeak}
\end{equation}
For rates with capacity subscriptions a new constraint is added to keep the load below a cap, $\alpha$:
\begin{equation}
    L \leq \alpha \text{.} \label{eq:cap}
\end{equation}
Where the total cost function is denoted by $f(L)$, this problem defines a linear program to optimally schedule the day's charging: 
\begin{equation}
\begin{aligned}
& \underset{r}{\text{minimize}}
& & f(L) \\
& \text{subject to}
& & \text{Eq} ~~\ref{eq:l},  \ref{eq:arr}, \ref{eq:dep}, \ref{eq:ene}, \ref{eq:rmax}, \ref{eq:rmin}. \label{eq:opt}
\end{aligned}
\end{equation}
In practice, uncertain driver behaviour necessitates more complex implementation using techniques from stochastic control or optimization under uncertainty. The aggregate profile resulting from those methods for implementation is similar to the ideal scenario; for our study we assume all parameters are known in advance.

\subsubsection{Machine Learning Model}

\begin{figure*}
    \centering
    \includegraphics[width=0.7\textwidth]{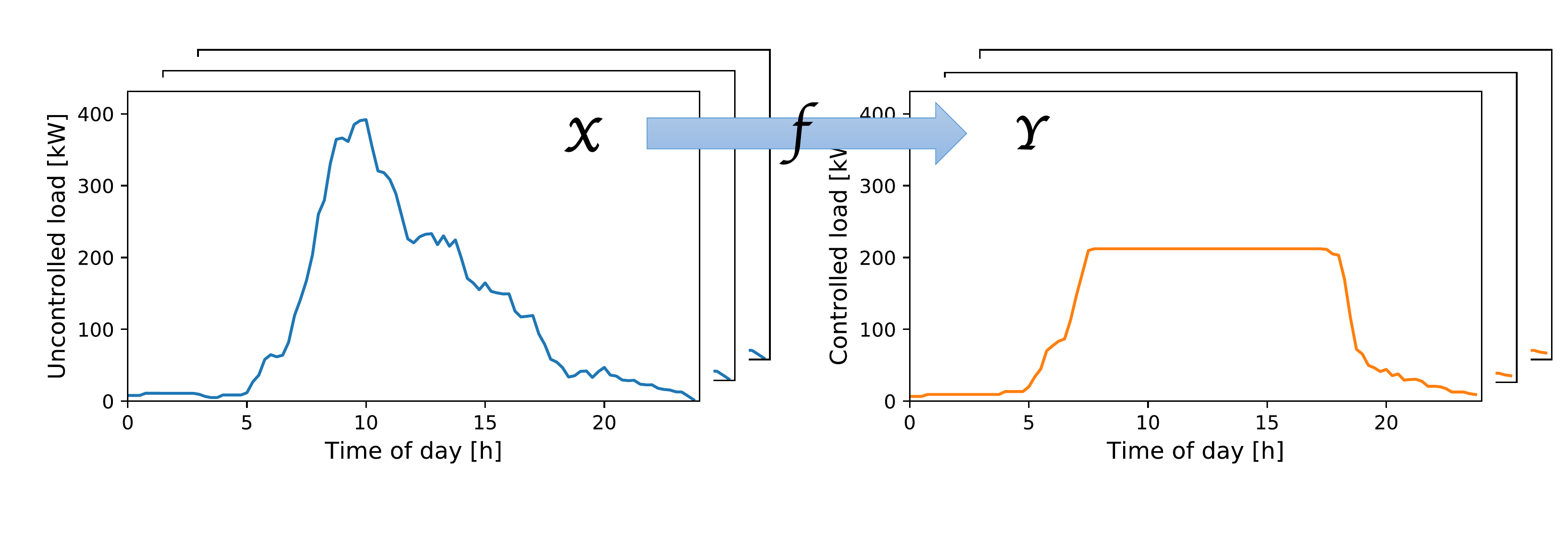}
    \caption{Framework for learning the control mapping, $f$. The training set depicted shows uncontrolled workplace charging profiles on the left for a parking lot with 250 vehicles. The corresponding controlled profiles on the right are calculated using peak minimization. }
    \label{fig:traintest}
\end{figure*}

The goal of this approach is to learn the best possible model for the mapping, $f$, shown in Figure \ref{fig:traintest}. To model the mapping we consider several options: linear regression, ridge regression, support vector regression with linear, quadratic, radial basis function, and sigmoid kernels, random forest regression, and shallow neural networks. Following best practices for machine learning, we divide the dataset into three parts: a training set (70\%), a development set (10\%), and a testing set (20\%). Models are trained on the training set and scores are reported for the testing set to show how the models perform on new samples \cite{james2013introduction}. 
The separate development set is used to tune some parameters of the pipeline, but the main fitting is done through cross-validation.

We implement cross-validation in the training set to train and tune the models because it gives a more robust score of performance than the single validation set. The procedure for five fold cross-validation: splits the training data into five parts, trains using four parts to report a score for performance on the fifth, and repeats that step five times holding out each part once. The mean score over the five folds is used to select the best parameters and the best mapping for the problem \cite{james2013introduction}. It is important to score the performance of each model on data that was not used for training, the fold left out in each step, to combat overfitting. 

All decisions are made using the cross-validation scores so that the score on the testing set can be viewed as an independent metric for performance on new samples.
This is implemented with scikit-learn \cite{pedregosa2011scikit}.

Linear regression models the mapping from input, $X_j$, to output, $Y_j$, as 
\begin{equation}
    \hat{Y_j} = A X_j + b \text{,}
\end{equation}
where $\hat{Y_j}$ is the model estimate for $Y_j$.
The coefficients, $A$ and $b$, are fit to minimize the residual sum of squares \cite{friedman2001elements}: 
\begin{equation}
    \min \sum_{j=1}^n \left(Y_j - \hat{Y_j}\right)^2 \text{.}
\end{equation}

Ridge regression is a variation on linear regression with added regularization in the form of a penalty proportional to the square of the L2 norm of the coefficients matrix \cite{james2013introduction}. The proportionality constant, $\alpha$, can be tuned.

Support vector regression is a kernel-based regression in the form:
\begin{equation}
    \hat{Y_j} = \sum_{i=1}^m \alpha_i K(X_i, X_j) + b \text{,}
\end{equation}
where $X_i$ are support vectors selected from the training data and $K$ represents the kernel function. In our study we consider four types of kernel: 
\begin{align}
    K(X_i, X_j) &= X_i^T X_j \\
    K(X_i, X_j) &= \left( \gamma X_i^T X_j + 1 \right)^2 \\
    K(X_i, X_j) &= \exp\left(- \gamma ||X_i-X_j||_2^2\right)^2 \\
    K(X_i, X_j) &= \text{tanh}\left( \gamma X_i^T X_j + 1 \right)
\end{align}
linear, quadratic, radial basis function (RBF), and sigmoid, respectively \cite{friedman2001elements}. The value $ \frac{1}{n_f}$ is used for the coefficient $\gamma$. Fitting the models requires tuning two parameters: $C$, for regularization, and $\epsilon$, to adjust the sensitivity of the loss function. We implement a grid search in the cross validation to find the optimal values for these parameters. 

Random forest regression is a method that fits and averages a large number of decision trees on subsets of the training data to achieve improved accuracy over using a single decision tree regressor \cite{friedman2001elements}. The method required tuning the number of trees and the maximum depth of each tree \cite{pedregosa2011scikit}.

Multi-layer Perceptron (MLP) regressors are a class of feed-forward neural networks \cite{friedman2001elements}. We limited the number of hidden layers to two and tuned the activation functions, the sizes of the hidden layers, and the regularization \cite{pedregosa2011scikit}.

For each sample, the uncontrolled load and the controlled load are both normalized by the maximum value of the uncontrolled load, $\max X_j$. This ensures the mapping, $f$, is independent of the number of cars in the aggregation and can capture both the shape change and the change in the magnitude of the peak.  

The different mappings are compared by their performance in Section \ref{sec:experiments} using the root mean squared error (RMSE) of estimates produced for the validation and test set:
\begin{equation}
    \text{RMSE} = \sqrt{\frac{1}{n}\sum_{j=1}^n \left(Y_j - \hat{Y_j}\right)^2} \text{.}
\end{equation}

The best model for each given rate schedule is selected, saved, and then applied to new uncontrolled charging profiles to estimate each scenario's corresponding controlled profile. The RMSE metric is in the same units as the samples, $Y_j$: since the samples are normalized, we can think of the RMSE as a percentage of the peak load.

Since $f$ is trained on normalized profiles, the procedure to apply $f$ is the same for scenarios with any number of vehicles: 1) normalize the uncontrolled aggregate charging load using its peak value, 2) apply the mapping, and 3) scale the result using the normalization constant to return the controlled aggregate charging profile. Applying this mapping is an inexpensive, closed-form calculation and the calculation cost is independent of the scenario scale.

\section{Data and Validation} \label{sec:experiments}
Our method is designed to work with any large dataset of charging sessions. In this paper we illustrate the method and present results using a dataset of charging from the California Bay Area described in Section \ref{sec:data}. 

\subsection{Dataset} \label{sec:data}
We use data from the operations of a major charging station provider, including all of their charging sessions in 9 California San Francisco Bay Area counties for the year 2019. That includes a total of 6.09 million sessions recorded for 119 thousand individual drivers. 4.2 million of the sessions are from workplace charging, 521 thousand are from L2 residential charging at single family homes, 148 thousand are from multifamily installations, and the remaining are from other public charging stations. No L1 charging is captured in the data.
The sessions include charging from both pure battery electric vehicles and plug-in hybrid electric vehicles.

\begin{figure*}
    \centering
    \includegraphics[width=\linewidth]{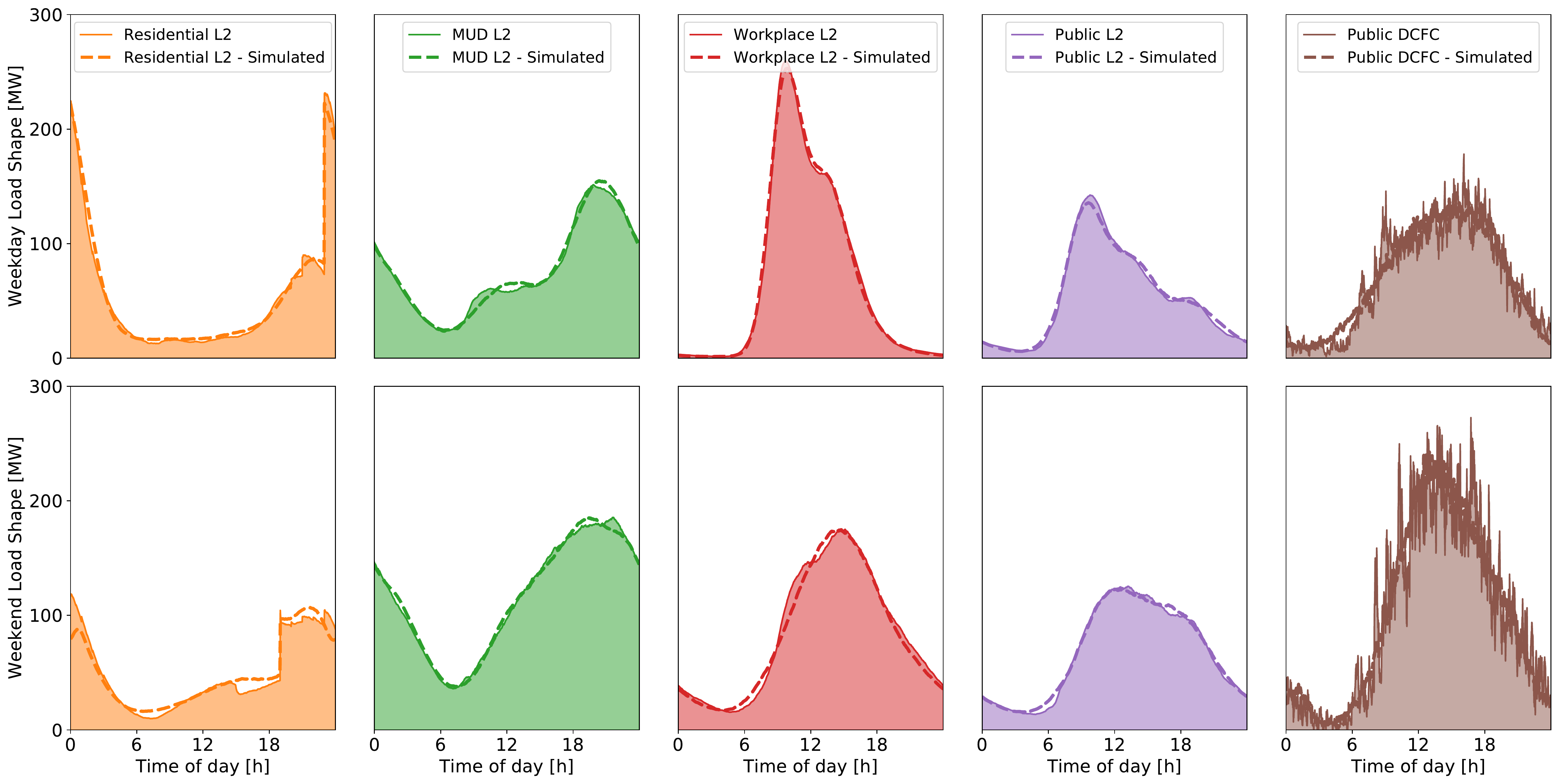}
    \caption{ Each subplot includes the load profile from 100,000 sessions. The solid lines and shading show profiles for which start times and energies were randomly sampled directly from the data, while the dashed line in each was calculated using start times and energies simulated by the Gaussian Mixture Models. 
    The top row shows weekday profiles and the bottom row shows weekend profiles. }
    \label{fig:load_samples}
\end{figure*}

Figure \ref{fig:load_samples} shows example large-scale load profiles for each segment. In these plots the public DCFC segment appears spiky because of its high charging rate, modeled as 150kW. The sessions are very short and often overlap: for example, when 100 of the 100,000 sessions have the same start time, a visible spike of 10.5 MW will occur. In the residential charging load, we can observe the impact of timers causing steep spikes in the evenings: the local utility has residential rate schedules for electric vehicle drivers under which the lowest price period starts at 11pm or 12am on weekdays and 7pm or 12am on weekends. 

\begin{figure}
    \centering
    \includegraphics[width=0.5\linewidth]{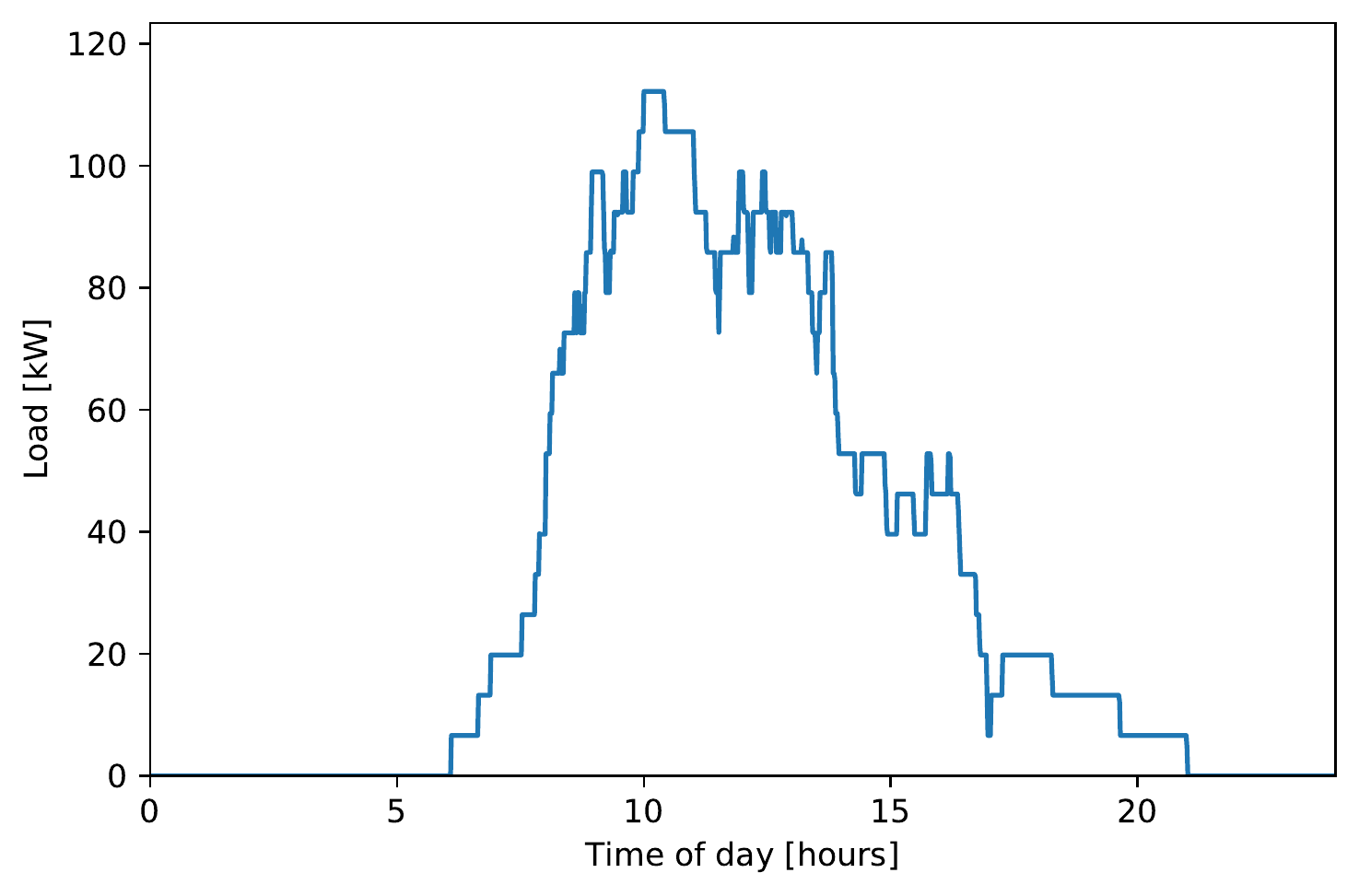}
    \caption{Illustration of a typical weekday load profile for a workplace parking lot with 50 sessions. }
    \label{fig:load_example}
\end{figure}

 To illustrate the load for an individual site, Figure \ref{fig:load_example} shows a sample workplace charging parking lot profile for 50 vehicles.

\subsection{Sessions Model Validation}

\begin{figure*}
    \centering
    \includegraphics[width=\linewidth]{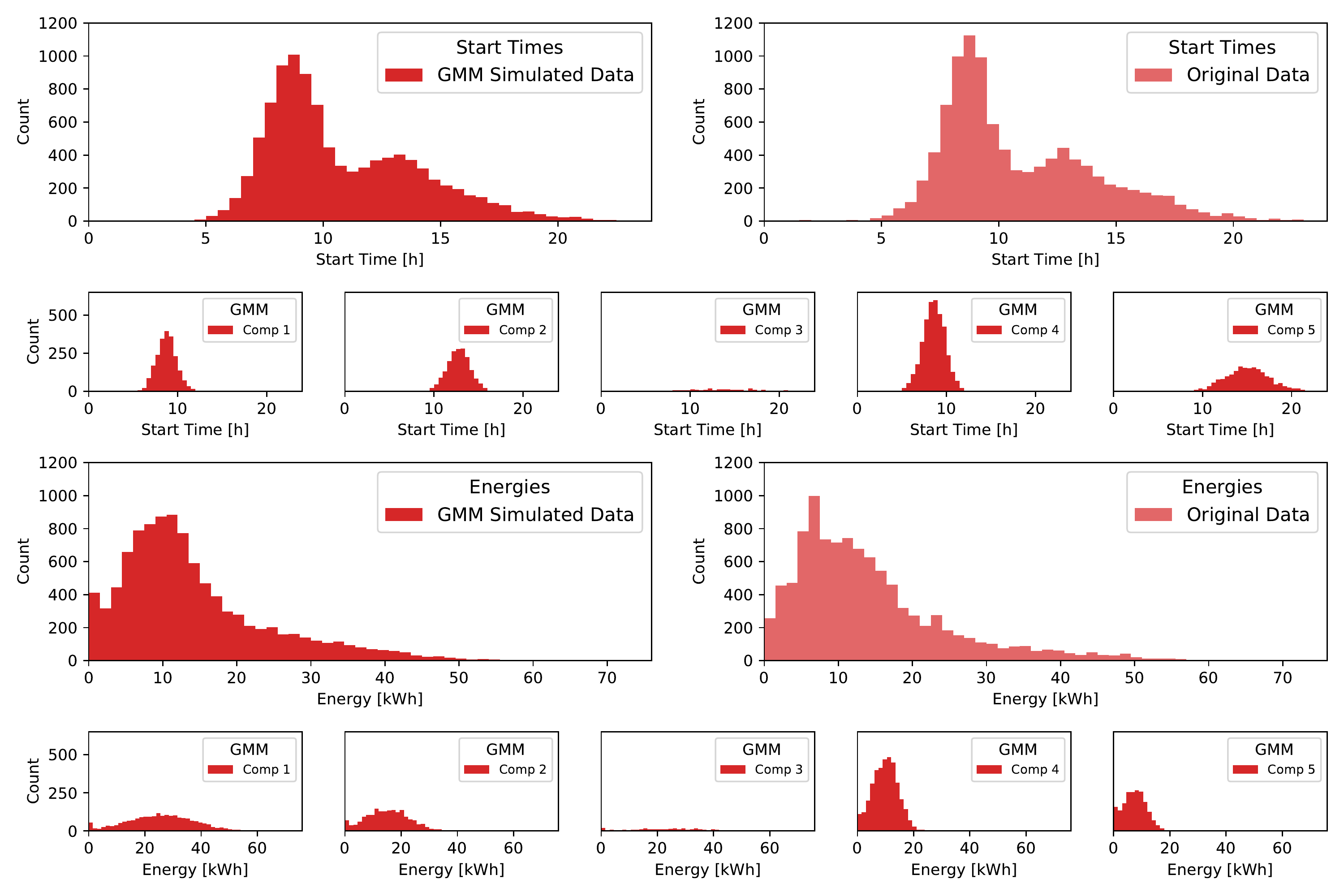}
    \caption{This figure shows the data and Gaussian Mixture Model (GMM) for weekday L2 workplace charging.
    The first two rows show the start time and the latter two rows show the session energy.
    In each part, the large histogram on the left shows simulated data generated by the GMM, the large histogram on the right shows values for a random sample of raw data, and the 5 smaller figures below illustrate the attribution of the simulated values to the GMM's components. }
    \label{fig:gmm_example}
\end{figure*}

A Gaussian Mixture Model (GMM) was fit to each segment in the data. Figure \ref{fig:gmm_example} shows weekday workplace charging as an example. 

For each model segment, the elbow curve of BIC was used to determine the optimal number of components. Of the 10 main segments (weekday and weekend; residential, multi-unit dwelling, workplace, public l2, public DCFC), we find 4 require 6 components, 3 require 5 components, 2 require 4 components, and 1 requires 8 components. This example in Figure \ref{fig:gmm_example} with workplace weekday charging requires 5 components in its mixture. To demonstrate that the mixture model accurately represents the distribution we compare sessions sampled from the raw data against sessions generated with the GMM.  
The results illustrate a good fit. The GMMs tend to smooth the distribution, and clipping the values for energy to be non-negative causes a small spike at zero; however, this has a small impact on the final results.

The load profiles in Figure \ref{fig:load_samples} present further validation of the model: dashed lines showing results calculated using the Gaussian Mixture Models match the original load profiles well. The simulated profiles are smoother than the raw data, especially for the segments for which we have less data, but the approach can capture the discontinuity caused by the residential timers well.

\begin{figure*}
    \centering
    \includegraphics[width=\linewidth]{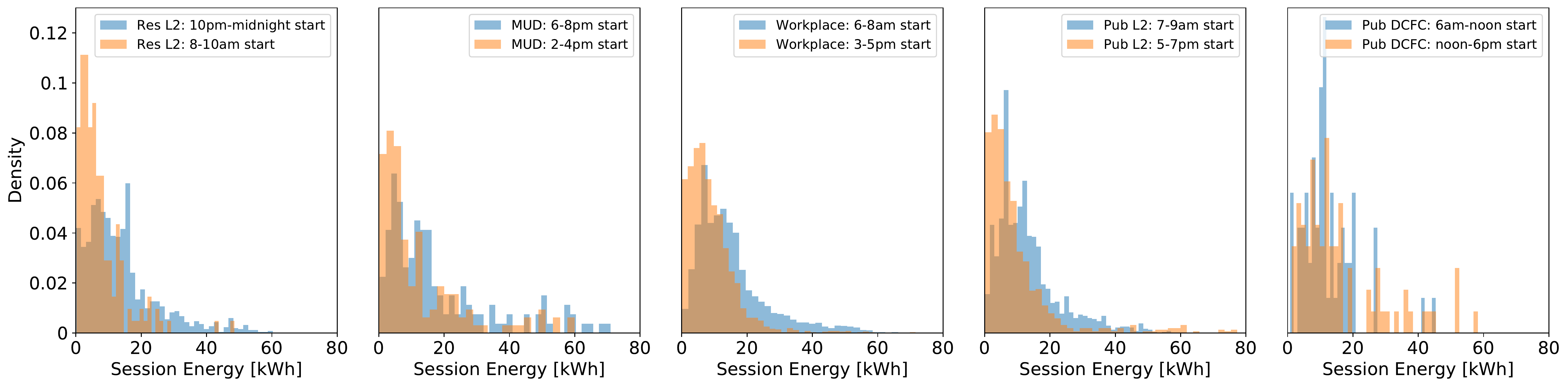}
    \caption{This figure shows how the sessions' start time window can impact the distribution of session energies, particularly in the workplace, public L2, and residential L2 segments. }
    \label{fig:joint}
\end{figure*}

We also validate the importance of modeling the joint distribution of start time and energy parameters, rather than modeling them separately. Figure \ref{fig:joint} illustrates how the distribution of session energies varies with the sessions' start times for different subsets and time windows across the charging segments. Within the three largest segments of the data, workplace, residential L2, and public L2, we can observe a clear difference. In the residential L2 segment, sessions starting in the morning are likely to require less energy than sessions starting at night. In the workplace segment, sessions starting in the afternoon, late in a typical workday, are likely to require less energy than those starting in the morning peak. In the public L2 segment, sessions starting in the morning are likely to require more energy. 

\subsection{Control Model Validation} \label{sec:control_results}

In these experiments we apply the model for load modulation control to workplace charging and we apply load shifting, or timer control, to residential charging. 

We implement workplace charging control for a set of representative rate schedules used by utilities in California for non-residential electric vehicle (EV) charging: first, Pacific Gas \& Electric's (PG\&E's) E19 rate schedule, which applies to medium size commercial buildings and often includes their parking lots; second, PG\&E's new commercial EV (CEV) rate schedule, which has a subscription set-up designed to encourage adoption; and third, Southern California Edison's (SCE's) TOU-EV-8 rate schedule for non-residential EV charging.

\begin{figure*}
    \centering
    \includegraphics[width=\linewidth]{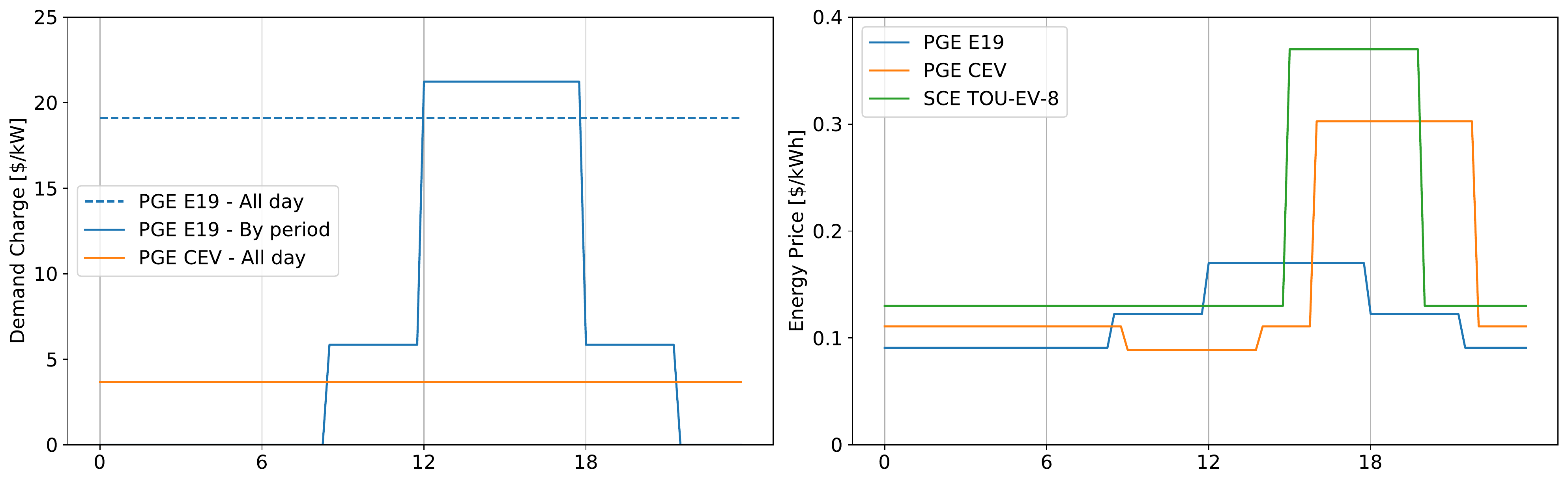}
    \caption{This figure shows the three utility rate schedules implemented in this study for workplace charging: PG\&E E19, PG\&E CEV, and SCE TOU-EV-8. Demand charge components are shown on the left; time of use pricing is shown on the right.  }
    \label{fig:all_rates}
\end{figure*}

These rates are composed of three parts: time-of-use (TOU) rates for energy, which give prices in dollars per kilowatt hour that vary throughout the day; demand charges, which give prices in dollars per kilowatt and apply to the peak total load reached during a given period of the day; and, unique to the PG\&E CEV rate, a subscription fee. All rate schemes and prices are presented in Figure \ref{fig:all_rates}. The mathematical representation of these components was discussed in Section \ref{sec:control}.

The SCE rate includes only TOU pricing. The PG\&E E19 rate includes TOU pricing varying throughout the day as well as demand charges for each the peak period, shoulder period, and and across the full day. The PG\&E E19 schedule's peak price period is during the afternoon, whereas the new PG\&E CEV schedules' peak period is in the evening. The PG\&E CEV rate includes a subscription fee wherein sites pay per 50 kilowatts to reserve capacity, and then pay only TOU energy costs for all consumption below that limit. We implement that subscription set-up in three ways: first, using an estimated capacity limit of 350 kW for our 250 vehicle sites, which forces a cap on the load but is weaker than a demand charge; second, using a proxy demand charge of the per-kW subscription price; and third, to study the impact of this rate's new TOU designs, using only the TOU rates with no limit or demand charges. 
We also implement a peak minimization scheme for comparison, which represents a hypothetical rate schedule thats only component is a full-day demand charge.

We create simulated parking lots of 250 vehicles, sampling randomly from the weekday workplace charging sessions in all counties to create each load profile. The size of the training and testing sets is important to limiting the complexity of the fitting. Through testing with the validation set on a subset of the models we found that no more than 1000 samples were required in the data set to achieve a minimal fitting error. 

We used a grid search with cross validation in the training set to find the best parameters for the tunable mappings. We report the results for each functional form on the testing set in Table \ref{tab:error_results}. The table uses the following short forms: LR denotes linear regression; RF denotes the random forest regressor; Ridge denotes linear ridge regression; Lin, Quad, RBF, and Sig SVR denote the support vector regressions with linear, quadratic, radial basis function, and sigmoid kernels, respectively; and MLP denotes the multi-layer perceptron regressor.

\begin{table*}
\resizebox{\textwidth}{!}{%
 \begin{tabular}{||c| c c c c c c c c||} 
 \hline
 Rate & LR & Ridge & RF & MLP & Lin SVR & Quad SVR & RBF SVR & Sig SVR  \\ [0.5ex] 
 \hline\hline
 Peak Minimization & \cellcolor{shade1!70} 0.02558 & \cellcolor{shade1!70} \textbf{0.02508} & \cellcolor{shade1!70} 0.02612 & \cellcolor{shade2!70} 0.02724 & \cellcolor{shade1!70} \textbf{0.02492} & \cellcolor{shade1!70} 0.02518 & \cellcolor{shade1!70} 0.02505 & \cellcolor{shade2!70} 0.02835 \\
 \hline
 PG\&E CEV & \cellcolor{shade7!70} 0.04854 & \cellcolor{shade6!70} 0.04748 & \cellcolor{shade6!70} \textbf{0.04554} & \cellcolor{shade6!70} 0.04667 & \cellcolor{shade7!70} 0.04808 & \cellcolor{shade7!70} 0.05130 & \cellcolor{shade7!70} 0.05144 & \cellcolor{shade8!70} 0.05632 \\
 \hline
 PG\&E CEV demand only & \cellcolor{shade1!70} 0.02365 & \cellcolor{shade1!70} \textbf{0.02305} & \cellcolor{shade1!70} 0.02375 & \cellcolor{shade1!70} 0.02630 & \cellcolor{shade1!70} \textbf{0.02288} & \cellcolor{shade1!70} 0.02324 & \cellcolor{shade1!70} 0.02303 & \cellcolor{shade2!70} 0.02743 \\
\hline
 PG\&E CEV energy only & \cellcolor{shade6!70} 0.04598 & \cellcolor{shade6!70} 0.04461 & \cellcolor{shade5!70} \textbf{0.04352} & \cellcolor{shade6!70} 0.04743 & \cellcolor{shade6!70} 0.04476 & \cellcolor{shade6!70} 0.04640 & \cellcolor{shade6!70} 0.04626 & \cellcolor{shade7!70} 0.05207 \\
 \hline
 PG\&E E19 & \cellcolor{shade2!70} 0.02889 & \cellcolor{shade2!70} 0.02828 & \cellcolor{shade2!70} \textbf{0.02775}& \cellcolor{shade2!70} 0.03109& \cellcolor{shade2!70} 0.02865& \cellcolor{shade3!70} 0.03140 & \cellcolor{shade3!70} 0.03128& \cellcolor{shade3!70} 0.03537 \\
 \hline 
 SCE TOU-EV-8 & \cellcolor{shade3!70} 0.03399 & \cellcolor{shade3!70} \textbf{0.03290} & \cellcolor{shade4!70} 0.03958 & \cellcolor{shade4!70} 0.03768	& \cellcolor{shade3!70} 0.03330 & \cellcolor{shade3!70} 0.03499 & \cellcolor{shade3!70} 0.03509	& \cellcolor{shade6!70} 0.04414\\
 \hline
\end{tabular}} 
\caption{Root mean squared error (RMSE) on the test set reported for each functional form and rate schedule. The shading divides the range of values into 8 colors, with darker reds showing the worse results. Bold font is used to highlight the best result for each rate based on performance in either the test set or cross-validation.} 
\label{tab:error_results}
\end{table*}

As discussed in Section \ref{sec:control}, the best model is selected based on its performance during cross-validation. In cross-validation, Ridge regression was the best performing model for the peak minimization, PG\&E CEV demand only, and SCE TOU-EV-8 rate schedule models. Random Forest regression was the best performing for the PG\&E CEV, PG\&E CEV energy only, and PG\&E E19 rate schedule models.
To present independent scores for performance on new samples, in Table \ref{tab:error_results} we present the results of applying each model to the reserved testing set. Because it is a different set of samples, those results are similar but have two slight differences from the cross-validation scoring: for both the peak minimization and the PG\&E CEV demand only implementations, Ridge regression performs best in cross-validation while Linear SVR performs best in testing. In cross-validation, for the first rate, Ridge regression returns a mean RMSE of 0.02590 and Linear SVR returns a slightly higher mean RMSE value of 0.02606; for the second rate, Ridge regression returns a mean RMSE of 0.02379 and Linear SVR returns 0.02406. The differences between the two are very small, at less than 0.03\% both in cross-validation and on the test set, so we are confident in the cross-validation result and continue with that selection.

We find that either ridge regression or random forest is selected for each of the different rate schedules. For many the two are very close: only for the PG\&E CEV cap and the PG\&E CEV energy-only rates does the random forest significantly outperform the ridge regression; only for the SCE rate does the ridge regression significantly out perform the random forest. Several of the other models also perform well, including, notably, linear regression and linear SVR. 

 We also observe that some rate schedules are easier to model than others, as the error values for rates with demand charges or for the peak minimization are significantly lower across the board. Variously, the CEV energy-only and other TOU rates seem to be harder to model.

\begin{figure*}
    \centering
    \includegraphics[width=\linewidth]{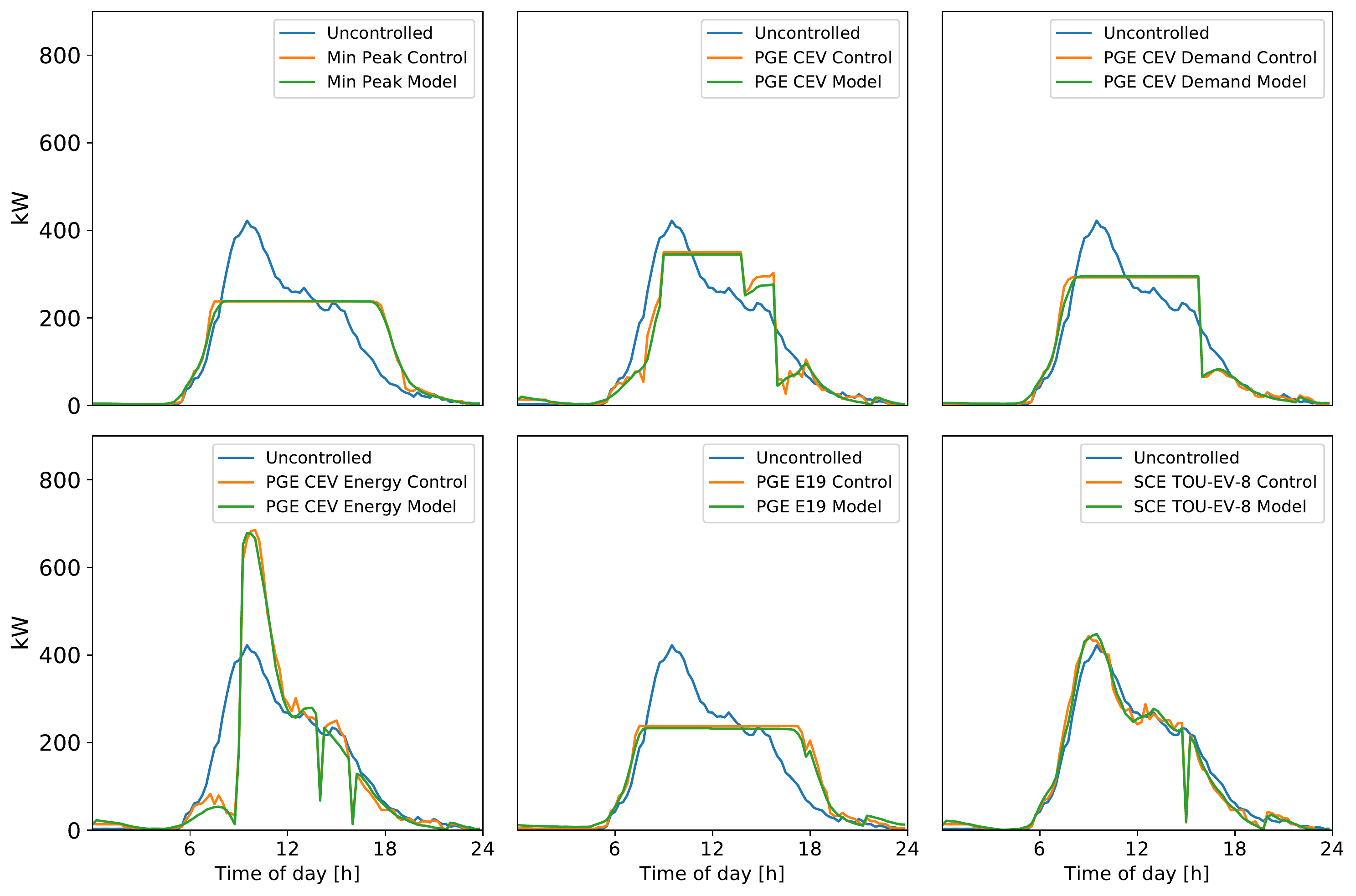}
    \caption{This figure illustrates the application of all 6 control schemes to a randomly sampled test profile drawn from the data set representing a workplace site with 250 vehicles. The `Model' series show the estimate of the best model for each control when applied to this sample. 
    The TOU-only rates have non-smooth behaviour near the transitions between prices because there is no regularization in the optimization implementation.}
    \label{fig:outputs}
\end{figure*}

Figure \ref{fig:outputs} shows performance of the best fit algorithm for each rate on a randomly sampled profile in the data set. A key observation from these results is the steepness in some controlled profiles. In the PG\&E CEV rates, the beginning of a new period in the rate can cause a dramatic and immediate drop or increase. Adding regularization to the control could smooth this effect at individual sites, but the drop is incentivized by the design of the rate structure with discontinuous pricing. 

\section{Results} \label{sec:results}
\subsection{Scenarios}
We present a set of results illustrating different scenarios for the future load from individual light-duty EV drivers at California's 2030 target of 5 million EVs \cite{brown2018}.

Each of these scenarios took approximately 42 seconds to run on a laptop computer. The same scenario with 1 million drivers, rather than 5 million, took 9.4 seconds to run on the same computer. Applying the control model to the workplace charging segment took an additional 8 seconds. 
The scenarios were modeled with 1 minute time intervals, except for those involving workplace charging control. The control models were fit using 15 minute time interval profiles to reduce computational costs and the corresponding scenarios which use those models are also presented here with 15 minute intervals. 

\subsubsection{Base Case}

Using the data set described in Section \ref{sec:data} we do not have direct measurements of L1 residential charging. To estimate the load from L1 residential charging we create a copy of the L2 residential mixture model and remove the components corresponding to the use of timers, reweighting the remaining components accordingly. This effectively returns a sessions distribution with only the uncontrolled residential charging behaviours. The `smooth' residential profiles shown in these scenarios were calculated using that altered mixture model. We assume there are no differences in the residential arrival times or energy requirements between L1 and L2 residential charging sessions: only the charging rate differs.

To create these scenarios we make assumptions about the split between segments: how many drivers in the 5 million depend on each type of charging? We assume each driver utilizes only one type of charging. To account for drivers' frequency of charging we make the additional assumption that residential type and workplace type drivers have an 80\% probability of charging each day, and public type drivers have a 33\% probability of charging each day. This lets us replicate the observed frequencies of charging while modeling only a single days profile. On weekends we assume the workplace type drivers probability of charging falls to 10\%. The assumption that each driver uses one type could be loosened to create other scenarios: the ultimate user inputs required are the number of sessions to simulate from each segment. 

Present day charging of individual EV drivers in California is heavily dominated by residential charging at single family homes, and a significant fraction of drivers use L1 \cite{wood2018california}. For these scenarios of 2030, however, we assume the situation will have developed: we assume that 10\% of the residential type drivers will be in the multi-unit dwelling segment, and we assume that only 20\% of the single family residential charging sessions will be L1. The remainder is assumed to be L2 charging and controllable through timers. 

\begin{figure*}
    \centering
    \includegraphics[width=\linewidth]{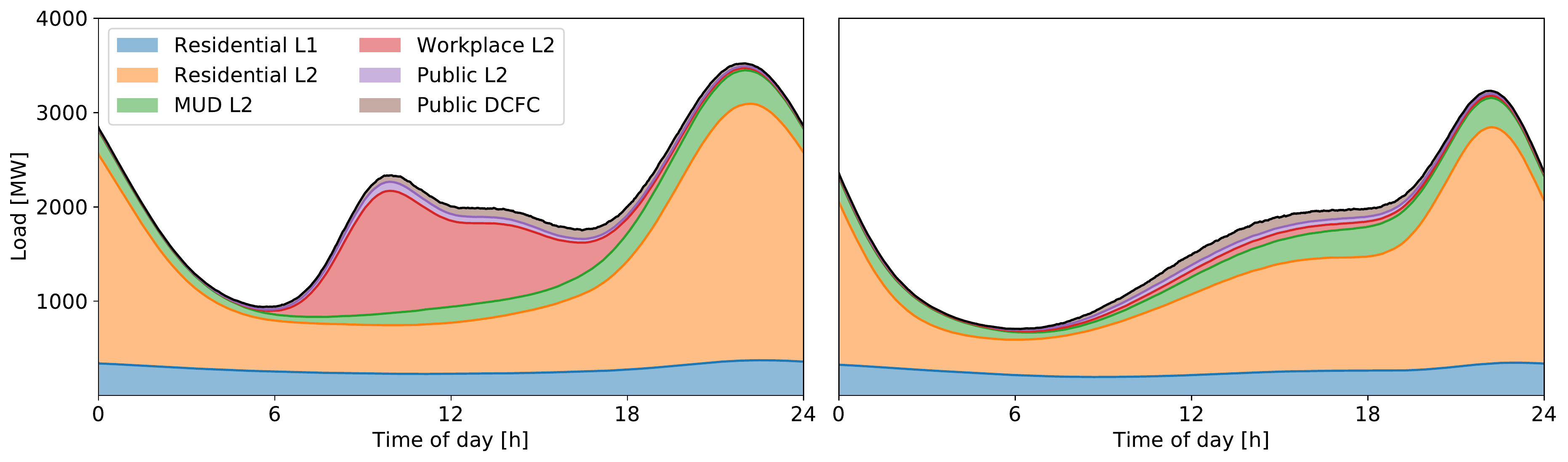}
    \caption{For a weekday on the left and a weekend day on the right, these show the base case scenario for 5 million EVs in 2030.}
    \label{fig:base_plots}
\end{figure*}

Figure \ref{fig:base_plots} shows a base case scenario, with 75\% residential drivers, 15\% workplace drivers, 5\% public L2 drivers, and 5\% public DCFC drivers. The peak load with these assumptions on a typical weekday is 3.522 GW occurring at 10 pm, and on a weekend is 3.232 GW occurring just after 10 pm. 

\subsubsection{Residential Timers}
\begin{figure*}
    \centering
    \includegraphics[width=\linewidth]{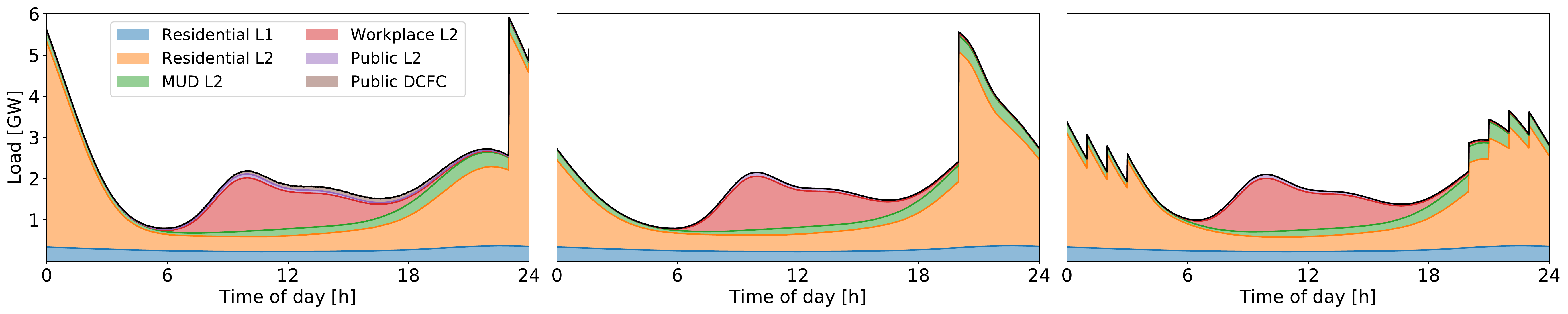}
    \caption{Adding timer control to the base case, investigating three scenarios with different participation levels and start times.}
    \label{fig:scen1}
\end{figure*}

We investigate changes to the residential charging load in the weekday scenario: first, by reintroducing the timer behaviour observed in today's data where TOU rates cause spikes at 11pm and 12am; second, by assuming a different rate schedule causes 25\% of the residential L2 type drivers to set timers for 8pm; and third, by assuming a new rate scheme causes 40\% of the residential L2 type drivers to set timers with start times staggered across every hour from 8pm to 3am (5\% at each). The results are shown in Figure \ref{fig:scen1}. 

We observe that including the timer behaviour significantly increases the peak in the first two scenarios and causes a sharp spike on the order of 3 GW at the start of the low price period. The third case, however, by staggering the start times has the effect of decreasing the peak relative to uncontrolled and successfully spreads the load more smoothly throughout the night. 

\subsubsection{Shifting Between Segments}
\begin{figure*}
    \centering
    \includegraphics[width=\linewidth]{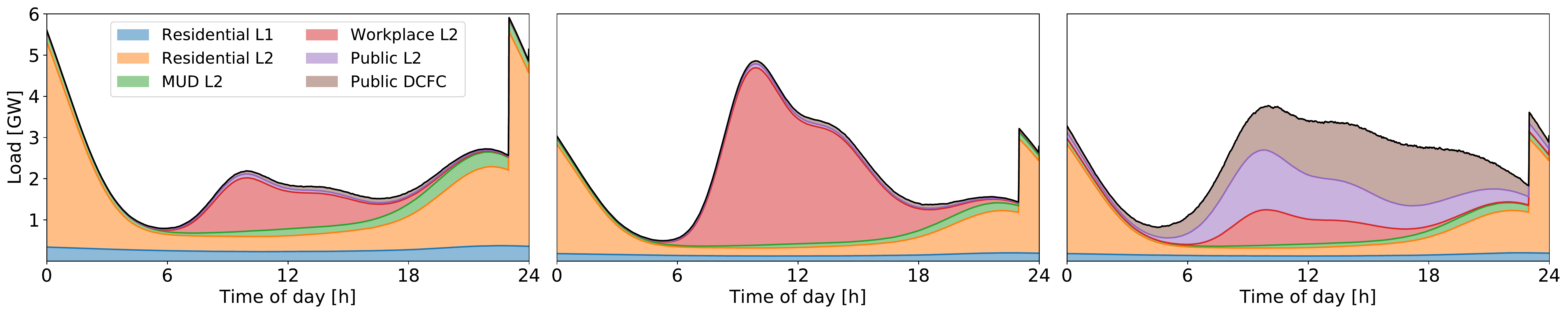}
    \caption{Comparing the division of drivers across segments in three scenarios (from left to right): residential, workplace, and public charging-focused. }
    \label{fig:scen2}
\end{figure*}

We also investigate the impact of distributing the drivers differently between the charging segments. Heavy investment in either workplace or public charging infrastructure, for example, could cause scenarios in 2030 where residential charging is much less common. Compared to a base case with 75\% residential drivers, how would the scenario change if workplace charging dominated? or if public charging, both fast and slow, became a dominant mode of charging? The results for these alternative scenarios are shown in Figure \ref{fig:scen2}. 
The first matches the base case from Figure \ref{fig:scen1}, with 75\% residential, 15\% workplace, 5\% public L2 and 5\% public DCFC; the second focused on workplace has only 40\% residential, then 50\% workplace, 5\% public L2, and 5\% public DCFC; and third focused on public has 40\% residential, 10\% workplace, and then 25\% public L2 and 25\% public DCFC. For the third case with high utilization of public charging we assume those drivers charge with 100\% probability on a given weekday. The residential L2 segment includes timers according to behaviour observed in today's data.

We observe a significant impact on the shape of the load: shifting drivers from residential to the public or workplace segments moves the load toward the middle of the day and the peak into the morning. 

\subsubsection{Workplace Charging Control}
We apply control to the workplace charging segment to investigate the impact of two key rate schedules. Starting with the workplace-dominant scenario presented in Figure \ref{fig:scen2} as a base case, we apply our models of PG\&E's E19 rate schedule, with TOU and demand charges, and  PG\&E's CEV rate schedule which includes a capacity limit. These controls are described in Section \ref{sec:control}. These scenarios assume that the whole workplace load is affected by the control, but the control could also be applied only to a fraction of the load. 

\begin{figure*}
    \centering
    \includegraphics[width=\linewidth]{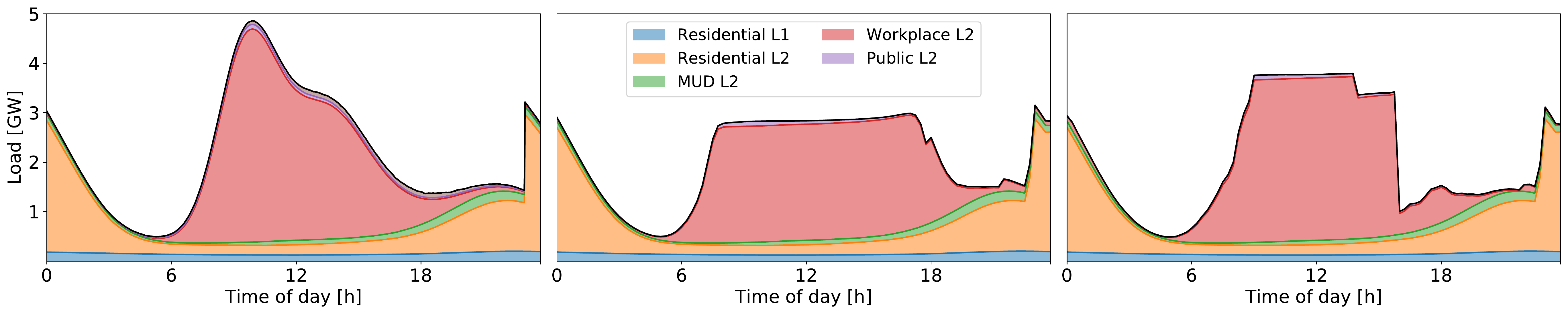}
    \caption{Compared to the uncontrolled workplace scenario from Figure \ref{fig:scen2}, this shows the impact of apply workplace charging control with two rate schedules: in the middle, PG\&E's E19, and on the right, PG\&E's CEV rate.}
    \label{fig:scen3}
\end{figure*}

The results presented in Figure \ref{fig:scen3} show that the control flattens the load relative to the uncontrolled profile and spreads it well throughout the workday, especially in the demand charge case with PG\&E's E19. Referring to Figure \ref{fig:outputs} we see that pure TOU rates can reshape the workplace load in other ways and even increase the peak if the lowest price period aligns with the morning peak of charging sessions, but each rate we studied that included a demand charge or limit on the peak load had this effect of flattening the load. 

The third case in Figure \ref{fig:scen3} with PG\&E's CEV rate shows a sharp drop at the beginning of the high price period around 4pm, matching the single site profiles studied in Figure \ref{fig:outputs}. This sharp drop, on the order of 3 GW, would be difficult to for the grid to sustain, similarly to the sharp spikes observed with the timers in Figure \ref{fig:scen1}.

\subsubsection{Summary}
These different scenarios show great variability in the load profile we can expect from 2030's fleet of 5 million personal EVs. 

In the base case of uncontrolled charging shown in Figure \ref{fig:base_plots}, we see the peak both on weekdays and weekends is aligned with the grid-wide evening peak typical in California; this is not the best-case scenario for the grid. In the residential charging investigation shown in Figure \ref{fig:scen1}, we see that widespread use of timers can either add to this problem if all drivers use the same setting by causing large spikes, or can help to flatten the load if start times are distributed. 

In Figure \ref{fig:scen2} we see that higher utilization of workplace and public charging can help shift the load towards the middle of the day, away from the evening peak and toward times with higher penetration of solar in the generation mix. The scenario dominated by public charging presents a flatter day-time load profile than the workplace charging scenario, but achieving that quantity of public DCFC may be expensive and pose challenges at the distribution level of the grid. 

In Figure \ref{fig:scen3} we see that the workplace load segment is very flexible and can be significantly redistributed through the day: with control, rates designed to optimize impact on this load profile, and investment to enable more public and workplace charging, this scenario could present an improvement in grid impact over the base case. This flattening also impacts the peak, and control with both rates shown in Figure \ref{fig:scen3} is found to significantly decrease the peak load relative to uncontrolled charging.

\subsection{Rate Design Application}
The steep spikes and drops observed in these scenarios could cause significant damage to grid infrastructure and pose challenges for grid operations. In California where the typical daily peak is on the order of 25 to 30 GW, a sudden drop or increase of 3 GW as seen in these scenarios represents a significant fraction \cite{caiso}. What rate design would improve the results presented in Figure \ref{fig:scen3} and improve the impact of EV charging on the grid? 

We propose rate design as an application of our model. For a given new rate schedule, what is the impact in a large-scale scenario on the load profile of EV charging? To answer this question for a rate implemented with load modulation controlled charging, we follow a simple procedure: 
\begin{enumerate}
    \item Generate new, artificial load profiles using the Gaussian Mixture Models for the segment,
    \item Build a training set for the control by implementing the optimization problem to minimize costs under the proposed rate schedule; repeat to create 1000 simulated instances,
    \item Fit a model to the optimization outputs,
    \item Generate a scenario of uncontrolled load,
    \item Apply the control model to the segment aggregate load profile to find the impact of the proposed rate in the scenario.
\end{enumerate}

The specific objective of a rate design problem depends on many inputs and is particular to each region; we propose this as a tool to help policy makers investigate the load impact as part of the design process. 

\begin{figure}
    \centering
    \includegraphics[width=\linewidth]{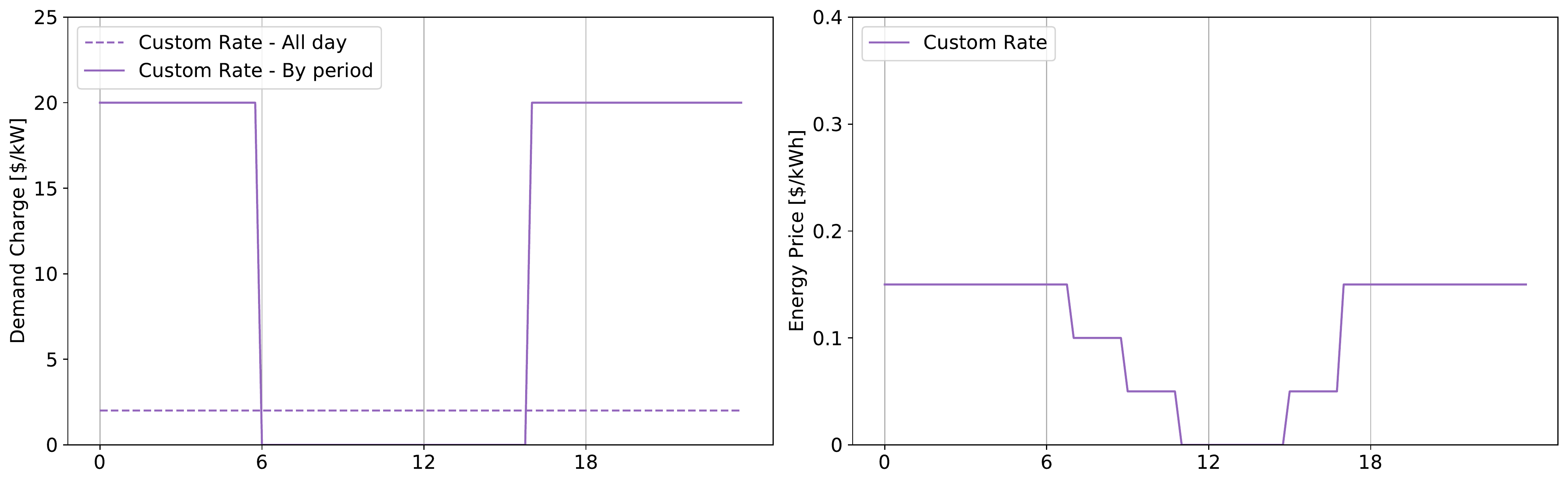}
    \caption{Illustration of the custom rate schedule: on the left, its demand charge components;  on the right, its TOU energy charge components. }
    \label{fig:custom_rate_definition}
\end{figure}

To demonstrate this procedure with an example we construct a new rate schedule for workplace charging which aims both to encourage charging during peak solar hours and to protect local grid infrastructure. The rate has heavy demand charge penalties for charging outside of solar hours and a moderate demand charge throughout the whole day to discourage peaks in the load. 
The details are presented in Figure \ref{fig:custom_rate_definition}. 

Executing the algorithm above with ridge regression to model the control, we find the model estimates match the true control optimization outputs well. The time break-down of the procedure illustrates the benefit of taking this approach. Run on a personal computer, it takes 45 minutes to prepare the training data by running the optimization for 1000 simulated parking lots of 250 vehicles, it takes under 3 seconds to fit a ridge regression model including a grid search over key parameters, and it takes under 50 seconds to build the scenario for all 5 million vehicles, 1 second of which is used to apply the control model to the workplace segment. 

By comparison, it would take over 5 hours to run the optimization in full for that scenario, dividing the 2 million workplace sessions into 8000 simulated parking lots for the calculation. 1 second marks a 18000 fold improvement over 5 hours.  If it were implemented as a single optimization problem, not dividing the sessions, the calculation would take even longer. Extrapolating from the time needed to optimize a 1000 car parking lot we estimate 2 million would take over 12 hours, but the calculation would likely fail due to the number of variables and the memory constraints of the laptop computer.

The 1 second calculation includes some time for pre- and post-processing the profile within the larger model framework. If we look at control for one individual parking lot of 250 vehicles: the optimization calculation takes 2.3 seconds; applying the pre-fit control model takes 0.0005 seconds. That marks a 4600 fold improvement. 

\begin{figure*}
    \centering
    \includegraphics[width=\linewidth]{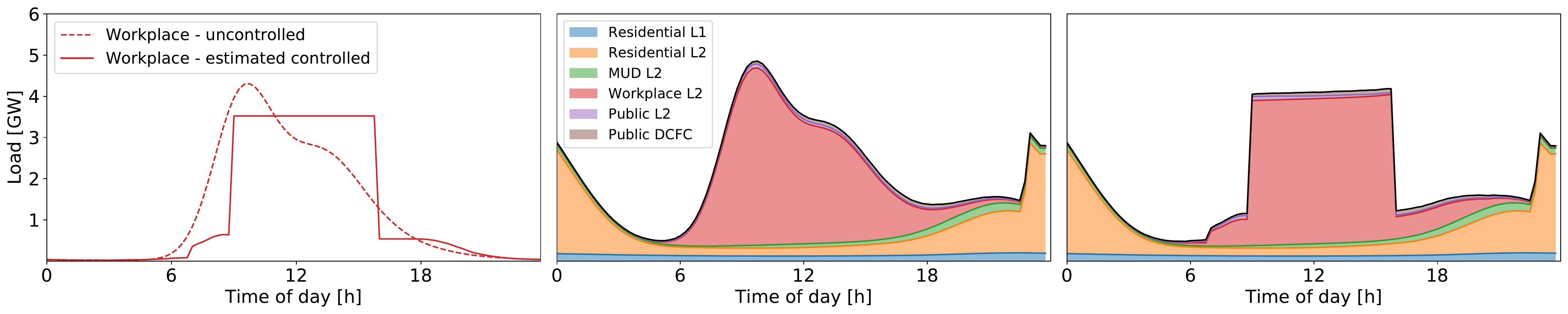}
    \caption{Implementation of a custom rate schedule: in the left subplot we show the workplace segment separately, in the middle subplot we present the uncontrolled base case (the same as in Figure \ref{fig:scen3}), and in the right subplot we combine the two to show the impact of the control within the overall scenario. }
    \label{fig:custom_rate}
\end{figure*}

In Figure \ref{fig:custom_rate} we show the impact on the workplace segment and on the overall scenario from Figure \ref{fig:scen3}. The results show that the two demand charge components drive the control, as there is no TOU-driven variation visible within the period of low demand charge during the day. That insight could help improve the rate design, suggesting a change in the relative strengths of the demand charge and energy price components. 

Overall this procedure is an inexpensive and fast tool to quantify the impact of workplace charging control under a new rate.

\section{Discussion} \label{sec:discussion}

The speed with which the model can be applied, both in generating scenarios and in applying different types of control, makes it a useful tool for many stakeholders with different applications in EV load modeling. Varying the assumptions to create different scenarios is possible in near real time. This can help a planner interacting with the model compare scenarios, understand the sensitivity of their results, and plan for a wider range of possible outcomes than if they were limited by the run-time of the tool. 

Since the modeling approaches separate working with the raw data from applying the model or generating scenarios, the model can be published and shared without raising privacy concerns for the individual drivers. Such privacy concerns have prevented many modelers from accessing real charging data; this approach, sharing statistical models of the data, could make charging scenarios more accessible. 

In addition to the design of the model, the results and applications demonstrated raise interesting insights about the future of rate design under automated charging control. 
Traditional rate design anticipates a range of user responses, not automated responses like the residential timers or software optimized workplace charging control, and our scenarios showed steep drops and spikes in the load resulting from the coordinated response of many vehicles to current rate structures. 

Our approach enabling study and anticipation of those optimized effects as part of the rate design process will helping rate designers and grid operators plan for the more optimized, automated future of charging.

\section{Conclusion} \label{sec:conclusion}

This paper proposed a modeling framework to assist long-term planning to support future charging demand from EVs, at and beyond the scale of today's ambitious EV targets. Building scenarios at the scale of 5 million or more EVs in under a minute enables a policy maker using the model to make quick comparisons, test assumptions, and interact with the model in near real-time. 

In this paper, we proposed a novel, data-driven approach to modeling load modulation charging control and demonstrated its application to a set of real electricity rate optimizations. This approach made it possible to include the impact of control in large-scale charging scenarios in real time: once the model was trained, applying each control rule to a new scenario added only a couple of seconds. We applied the framework to a set of scenarios representing California's 2030 demand for EV charging, highlighting the impact of changing assumptions between scenarios and applying control to the charging. We also demonstrated how the framework can be used to study new rate schedules: using sessions data simulated with the sessions model and fitting a new control model using our machine learning approach, we quickly estimated the large-scale impact of a new workplace electricity rate schedule. 

This approach introduces a new privacy-preserving tool for planners and policy makers to generate rapid, data-driven, scalable scenarios for EV charging and study the impact of different charging controls, designed for the scale of adoption we will need to meet the electrification targets set today by countries around the world. The model has been published in an open-source tool, available at \url{https://github.com/slacgismo/SCRIPT-tool/} \cite{cezar2021}, along with the sessions and control model objects learned for the scenarios in this paper.

\subsection{Future Work}

The charging optimization behind the control model is designed for an ideal scenario where all information is known: incorporating uncertainty and adding regularization to smooth the spikes caused by TOU control would make the control model more similar to real implementations. Future work on the load modulation control model could also focus on developing a theoretical understanding for why particular methods perform better for particular styles of rate design. If the model can be parameterized directly on the rate, that will enable a closed-loop tool for optimizing the rate design process. 

The model's open-loop design can be a limitation. Changing the proportion of drivers charging in each segment means sampling a different number of sessions from each distribution, not selecting particular users and altering their charging sessions. Since each distribution has different properties, in particular for session energy, this means the total energy between scenarios is not held constant. To alter the approach would require many assumptions about which of the drivers switch or stay across scenarios.

There is information about driving patterns and needs baked into the different energy distributions. Further investigation is necessary to discover whether segments in the data with higher average session energies reflect use by drivers who charge less frequently, or whether access to those segments encourages drivers with higher daily mileage to electrify. A future version of the model could consider which drivers will electrify by 2030 to determine their driving needs and their likelihood to charge in each segment in greater detail. 

\section{Acknowledgements}

The authors would like to thank their colleagues in the Stanford Sustainable Systems Lab (S3L) at Stanford University led by Prof. Ram Rajagopal and the Grid Integration Systems Mobility group (GISMo) at SLAC National Accelerator Laboratory for their continued support. The authors would also like to thank their colleagues at Energy and Environmental Economics (E3), Prof. Mahnoosh Alizadeh at the University of California Santa Barbara, ChargePoint, Gridmatic and the California Energy Commission staff, in particular Reynaldo Gonzalez, that made this work possible.

This work was funded by the California Energy Commission under grant EPC-16-057 and by the National Science Foundation through a CAREER award (\#1554178). SLAC National Accelerator Laboratory is operated for the US Department of Energy by Stanford University under Contract DE-AC02-76SF00515.

\section{Declaration of Interest}
All work presented in this paper was conducted in accordance with Stanford University's Conflict of Interest Policies.

\end{document}